\documentclass[twocolumn]{aastex6}
\usepackage{subcaption}
\usepackage{amsmath,amssymb,amstext}
\usepackage{natbib}
\usepackage{xspace}

\newcommand{\HEALPix}{\texttt{HEALPix}\xspace}
\newcommand{\LIGO}{LIGO\xspace}
\newcommand{\VIRGO}{VIRGO\xspace}
\newcommand{\Fermi}{\emph{Fermi}\xspace}

\newcommand{\Swift}{\emph{Swift}\xspace}
\newcommand{\INTEGRAL}{\emph{INTEGRAL}\xspace}

\newcommand{\red}{}
\newcommand{\blue}{}
\newcommand{\green}[1]{\textcolor{blue}{#1}}

\begin{document}
\title{Searching for High-Energy \boldmath$\gamma$-Ray Counterparts to Gravitational Wave Sources with \Fermi-LAT: a Needle in a Haystack}
\author{G.~Vianello\altaffilmark{1}}
\author{N. Omodei\altaffilmark{1}}
\author{J.~Chiang\altaffilmark{1}}
\author{\blue{S.~Digel}\altaffilmark{1}}
\altaffiltext{1}{W. W. Hansen Experimental Physics Laboratory, Kavli Institute for Particle Astrophysics and Cosmology, Department of Physics and SLAC National Accelerator Laboratory, Stanford University, Stanford, CA 94305, USA}

\begin{abstract}
At least a fraction of Gravitational Wave (GW) progenitors are expected to emit an electromagnetic (EM) signal in the form of a short gamma-ray burst (sGRB). Discovering such a transient EM counterpart is challenging because the \LIGO/\VIRGO localization region is much larger (several hundreds of square degrees) than the field of view of X-ray, optical and radio telescopes.
The \Fermi Large Area Telescope (LAT) has a wide field of view ($\sim 2.4$ sr), and detects $\sim 2-3$ sGRBs per year above 100 MeV. It can detect them not only during the short prompt phase, but also during their long-lasting high-energy afterglow phase. If other wide-field high-energy instruments such as \Fermi-GBM, \Swift-BAT or \INTEGRAL-ISGRI cannot detect or localize with enough precision an EM counterpart during the prompt phase, the LAT can potentially pinpoint it with $\lesssim 10$ arcmin accuracy during the afterglow phase. This routinely happens with gamma-ray bursts. Moreover, the LAT will cover the entire localization region within hours of any triggers during normal operations, allowing the $\gamma$-ray flux of any EM counterpart to be measured or constrained. We illustrate two new \emph{ad hoc} methods to search for EM counterparts with the LAT, and their application to the GW candidate LVT151012.
\end{abstract}
\maketitle

\section{Introduction}

The first direct detection of a Gravitational Wave event (GW) by the recently upgraded \LIGO \citep{1992Sci...256..325A,2009PhRvD..80j2001A, 2016PhRvL.116f1102A} opened a new era in astronomy. 
The first science run `O1' with the Advanced \LIGO detector started in September 2015, and two high-significance events (GW150914 and GW151226) and one sub-threshold event (LVT151012) were reported \citep{2016PhRvL.116f1102A, 2016PhRvL.116x1103A}. 
These three events were compatible with the signal expected from the merger of two Black Holes (BH). 
In future \LIGO science runs, additional sources of GW events might include the mergers of other compact object binary systems: neutron star-black hole (NS-BH) and neutron star-neutron star (NS-NS).

The identification and study of electromagnetic counterparts (EM) to GW events is important for several reasons. It resolves degeneracies associated with the inferred binary parameters and allows for a cross-check between the distances measured through the GW signal with the redshifts measured through the EM counterpart, providing an independent constraint on cosmological models. The simultaneous detection of a clear EM counterpart can also confirm near-threshold or sub-threshold GW events, effectively increasing the sensitivity of the search and the distance to which GW events can be detected by \LIGO/\VIRGO. The potential wealth of complementary information encoded in the EM signal is likewise essential to fully unravel the astrophysical context of the event. However, discovering an EM counterpart is challenging because localization regions of GW event provided by \LIGO/\VIRGO are currently as large as several hundred square degrees, much larger than the fields of view (FoVs) of typical X-ray, optical or radio telescopes \citep{2016arXiv160208492A}. The luminosity of the expected EM counterpart is also expected to decay rapidly, leaving a short time window to complete the coverage of the localization region. On the other hand, hard X-ray telescopes such as \Swift-BAT \citep{2005SSRv..120..143B} and \INTEGRAL-ISGRI \citep{2003A&A...411L.291M}, as well as $\gamma$-ray detectors such as the \Fermi Gamma-Ray Burst Monitor \citep[GBM,][]{2009ApJ...702..791M} \Fermi Large Area Telescope \citep[LAT,][]{2009ApJ...697.1071A} and HAWC \citep{2012APh....35..641A}, have much larger FoVs and can cover the localization region much more quickly. They are therefore expected to play a major role in the discovery of the first EM counterpart to a GW event.

Short gamma-ray bursts (sGRBs) are a class of GRB with durations $\lesssim 2$ s, and they are thought to be associated with the mergers of BH-NS or NS-NS binaries (\citealt{Eichler1989}, \citealt{Narayan1992}, \citealt{LeeRamirezRuiz2007}, and \citealt{Nakar2007}). They are therefore the expected EM counterparts for GW events involving at least one NS. \Fermi-GBM is the most prolific detector of sGRBs ($\sim 40$ per year), and it is likely to be the first instrument to firmly detect an EM counterpart to a GW event \citep{GW150914_GBM}. However, it localizes sGRBs with uncertainties of the order of a few degrees, making the follow-up by instruments at other wavelengths challenging.
Coded-mask telescopes such as BAT and ISGRI can localize events with arcmin precision, but they have smaller FoVs and indeed detect $\sim 8$ sGRBs per year. HAWC has a very large FoV, but as yet has not detected its first GRB. 

The \Fermi observatory was launched in June 2008 and orbits the Earth at an altitude of $\sim 560$ km with a period of 96.5 minutes. The \Fermi-LAT is a pair-conversion telescope that detects $\gamma$ rays in the energy range from 20\,MeV to more than 300\,GeV. It has a FoV of $\sim 2.4$ sr and it covers the entire sky every $\sim 3$ hours during normal operations. 
It detects around 15 GRBs per year, among them 2--3 are of the short-duration class, with localization of the order of $\sim 10$ arcmin \citep{2015arXiv150203122V}. When detected by the LAT at high energy ($>$ 100 MeV), sGRBs have a substantially longer duration with respect to their keV--MeV emission. This long-lasting emission is thought to be related to the so-called afterglow emission, also observed at other wavelengths \citep{grb090510,grb110731,kouveliotou13}. \Fermi-LAT is the only wide-field instrument that has detected and localized an sGRB \textit{during its afterglow phase} starting from the GBM localization.  MASTER and iPTF have been able to do the same, but only for \textit{long} GRBs so far \citet{2016MNRAS.455..712L, 2015ApJ...806...52S}. Should the detection of an EM counterpart be made by the GBM, \Fermi-LAT could substantially reduce the localization uncertainty, facilitating follow-up at other wavelengths. Should the counterpart be occulted by the Earth for the GBM, and outside the FoV of coded mask instruments, then the LAT would be the only instrument that could still detect the GRB in the $1$--$2$ hours after the burst. Therefore, \Fermi-LAT plays a unique role in facilitating the multi-wavelength follow-up of GW events. This happens routinely already for GRBs, as the vast majority of GRBs detected by the GBM that were also localized by the LAT, were then successfully followed up by other instruments.  Among other results, this led to the spectroscopic measurements of more than 20 redshifts \citep{2015arXiv150203122V} for GRBs.

Very bright GRBs can also be localized by the LAT on-board with an accuracy between $0.1$ and $\sim 0.5$ deg. This localization is then distributed within seconds to observatories on the ground, allowing for quick follow-up. This  happened four times during the first 8 years of the \Fermi mission \citep{2009GCN..9334....1O,2013GCN..15464...1R,2016GCN..19403....1O,2016GCN..19580....1O}.


\blue{BH-BH mergers are sources of GW as well, but are not expected to produce EM signals. Nonetheless the \Fermi-GBM observation of a low-significance candidate counterpart 0.4 s after GW 150914 warrants searches for counterparts for GW produced by this class of progenitors.}\footnote{We note that \citet{Lyutikov2016} contests this association on the grounds of the constraints it imposes on the circum-merger environment, while other authors argue against it mainly due to the non-detection by the \INTEGRAL-ACS instrument \citep{greiner16,GW150914_integral}.}
The possible association between BH-BH mergers and $\gamma$-ray transients will be addressed by future GW events. If confirmed, it would constitute a surprise that would require new models, such as those published following the report of GW150914-GBM \citep[e.g.][]{Loeb2016, Fraschetti2016, Janiuk2016, 2016ApJ...821L..18P}. 
Some of these new models foresee a counterpart similar to a standard sGRB, that would imply a possible afterglow signal in the LAT.


The standard LAT analysis assumes the source location to be known with some accuracy.
Given the size of the localization region of a GW event, the search for a transient counterpart in LAT data is challenging and requires \emph{ad-hoc} methods. In the case of a non-detection, constraining the flux of the EM counterpart requires accounting for the uncertainty on the position of the source, which requires a careful statistical treatment. In this paper, we detail two new methods to search for EM counterparts to GW events in \Fermi-LAT data and to constrain their fluxes. A comprehensive presentation of the results of the \Fermi-LAT follow up for three GW events using the methods presented here is provided in \citet{GW150914_LAT} and \citet{Racusin16}.

\section{Searching for EM counterparts to GW events in \Fermi-LAT data}

Compact Binary Coalescence events discovered by the \LIGO and \VIRGO experiments (\citealt{2016PhRvL.116f1102A}, \citealt{2016PhRvL.116x1103A}) are localized with a Bayesian analysis that uses nested sampling to produced a marginalized posterior probability for the localization of the transient. This localization is distributed by the \LIGO/\VIRGO collaboration in the form of a full-sky map in \HEALPix format \citep{HEALPix}, an equal-area projection, for which the value of each pixel $p_{h}$ is the integral of the probability density over that pixel, so that
\begin{equation}
\sum_{h=0}^{H} p_{h} = 1,
\end{equation}
where $H$ is the number of pixels.

\begin{figure}[tb]
\centering
\includegraphics[width=0.48\textwidth,trim=190 80 190 50,clip=true]{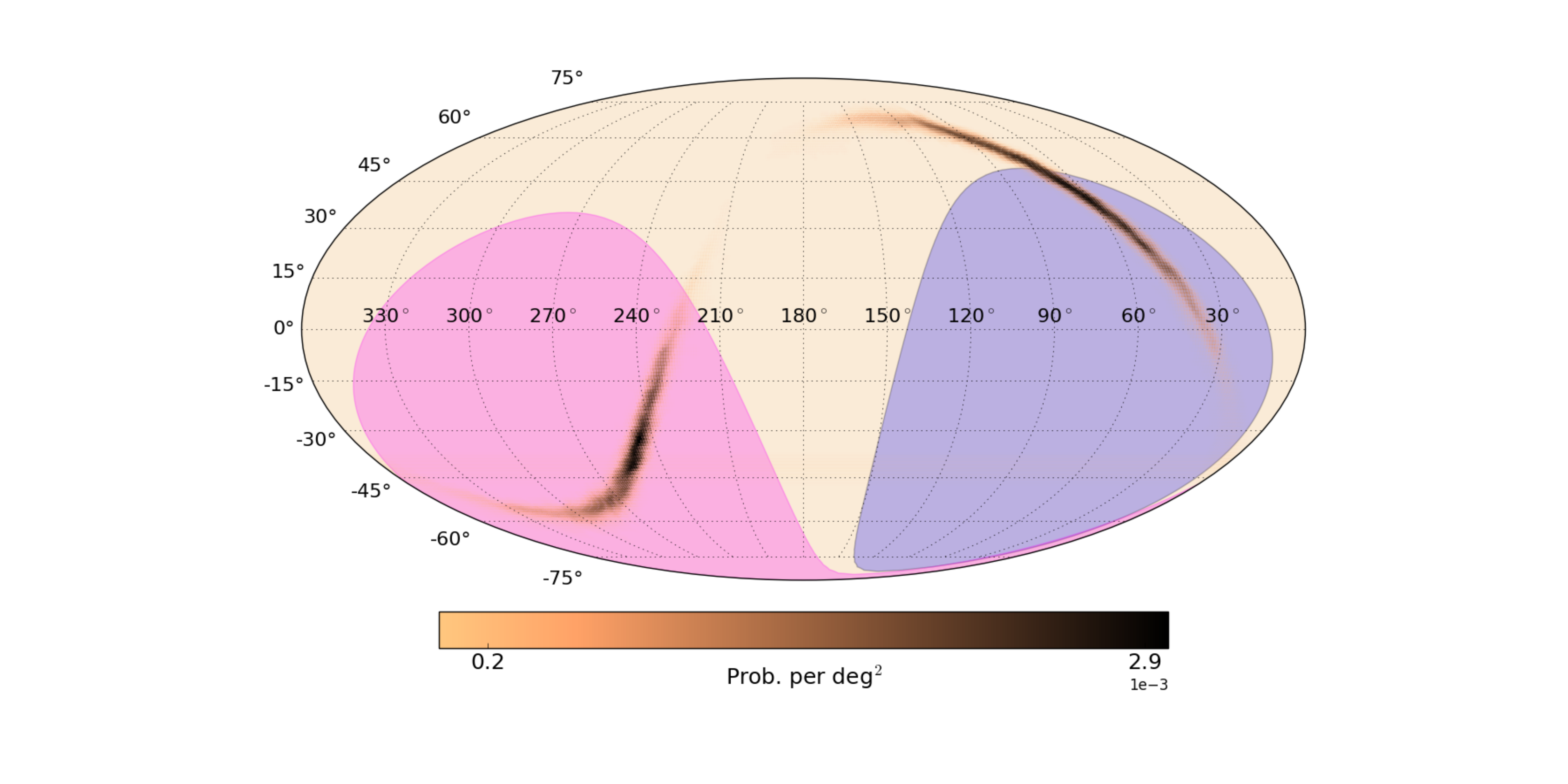}
\caption{\LIGO localization probability maps for LVT151012. The blue region represents the part of the sky occulted by the Earth at the trigger time from the vantage position of \Fermi, $\sim 560$ km above the Earth surface. The pink region represents the LAT field of view at the time of the trigger \citep[reproduced from][]{Racusin16}.} \label{fig:ligomaps}
\end{figure}

In Fig.~\ref{fig:ligomaps} we show \blue{the probability map} for LVT151012 \citep{2016PhRvL.116x1103A}. The size of the 90\% containment region for such a localization is usually of the order of several hundreds of square degrees. This will be reduced as more GW detectors will come online. The search for a transient counterpart in such a large portion of the sky requires an \emph{ad-hoc} treatment, described in the following sections.

\subsection{LAT analysis}
\label{sec:lat_analysis}

All of our analyses are based on the standard unbinned maximum likelihood technique used for LAT data analysis\footnote{\url{http://fermi.gsfc.nasa.gov/ssc/data/analysis/documentation/Cicerone}} \green{based on Poisson statistic}.
We include in our \textit{baseline} likelihood model all sources (point-like and extended) from the LAT source catalog (3FGL, \citealt{3fgl}), as well as the Galactic and isotropic diffuse templates provided by the \Fermi-LAT Collaboration \citep{2016ApJS..223...26A}.  
When needed, we employ a likelihood-ratio test \citep{Neyman1928} to quantify whether the existence of a new source is statistically warranted. In doing so, we form a test statistic (TS) that is twice the logarithm of the ratio of the  likelihood $L_{1}$ evaluated at the best-fit model parameters when including a candidate point source at a given position (alternative hypothesis) to the likelihood $L_{0}$ evaluated at the best-fit parameters under the baseline model (null hypothesis):
\blue{\begin{equation}
{\rm TS} = 2~(\log{L_{1}} - \log{L_{0}}).
\end{equation}
\green{When comparing a background model and a background model plus a point source at a fixed position as in our case, TS is distributed with good approximation as $\frac{1}{2}\chi^{2}$ with 1 degree of freedom. This result has been obtained with Monte Carlo simulations in \citet{1996ApJ...461..396M}. The factor $1/2$ comes from the fact that we force the point source to have a positive flux. Therefore, for one test, the significance of an excess can be estimated as $\sigma=\sqrt{TS}$.}
For all the likelihood analyses described in the next sections, we use the Pass 8 \texttt{P8\_TRANSIENT010E\_V6} event class and the corresponding instrument response functions.

In the first method described in this paper (section~\ref{sec:fixed-time-search}), we look for a possible EM counterpart using a standard frequentist technique, namely the Maximum Likelihood (ML) analysis commonly used for \Fermi-LAT analysis\footnote{\url{http://fermi.gsfc.nasa.gov/ssc/data/analysis/documentation/Cicerone}}. In case of non-detection, we choose instead a Bayesian technique to constrain the flux $F$ of the EM counterpart. The Bayesian framework provides a natural way to account for the information contained in the LIGO probability map, which constitutes a prior for our analysis, and allows us to define a ``global'' upper bound $F_{ub}$ for $F$ despite the uncertainty in the localization of the GW signal, as well as the presence of nuisance parameters. Following \citet{2015NIMPA.774..103B} we note that $F_{ub}$ can be used directly to place constraints on theoretical models.

In the second method \blue{(section~\ref{sec:adaptive-time-search})} we do not use directly the information contained in the LIGO probability map, except for limiting the search to the 90\% containment region. We again use ML for the search for a counterpart. This time we follow the method of \citet{1983NIMPR.212..319H} to measure an upper bound which expresses a constraint on the flux of the source in case of non-detection, profiling out all nuisance parameters. This is a standard method implemented in the \textit{Fermi Science Tools}. 
In section~\ref{sec:trials} we describe how we deal with the multiple likelihood-ratio tests involved in our searches. We then provide our conclusions in section~\ref{sec:disc}.


\subsection{Fixed time window search}
\label{sec:fixed-time-search}

\begin{figure}[tb]
\centering
\includegraphics[width=0.48\textwidth]{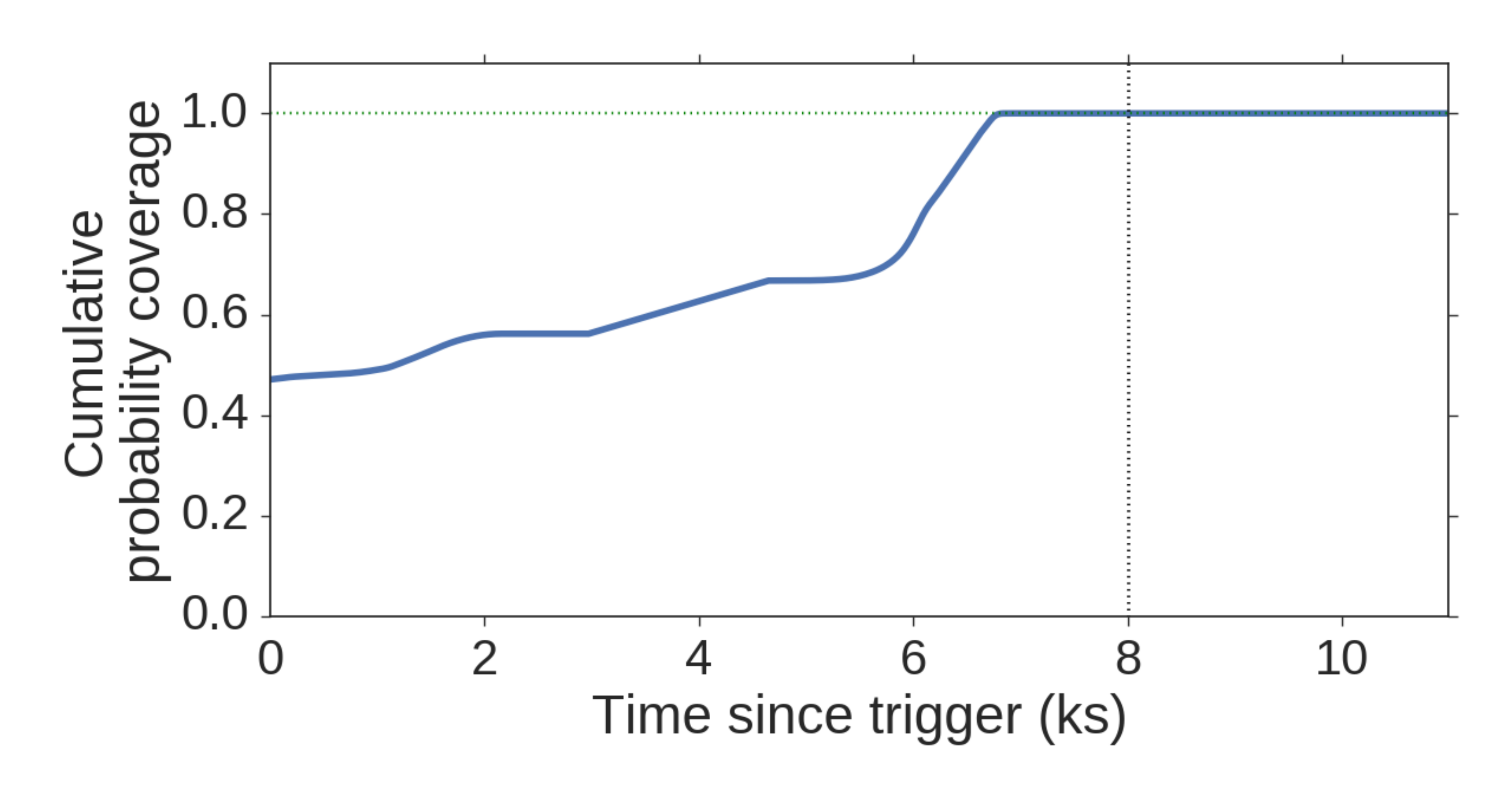}
\caption{Cumulative sum of the probability of all the pixels observed within a given time since the trigger LVT151012 (see text for details). The vertical dashed line indicates the end of the time window used for the fixed time window analysis.} \label{fig:coverage}
\end{figure}

In this analysis we search for high-energy $\gamma$-ray emission on a fixed time window that starts at the time of the \LIGO trigger. Since \Fermi changes attitude continuously, covering the full sky in just under 3 hours, different points in the LIGO localization region are observed at different times. An obvious choice for the time window is then the shortest time interval where the LIGO localization has been covered completely. For the analysis presented in this paper, a direction in the sky is considered observable by the LAT if it is within  65$^\circ$ of the LAT boresight (or LAT z-axis) and has an angle with respect to the local zenith smaller than 100$^\circ$. The latter requirement is used to exclude contamination from terrestrial $\gamma$ rays produced by interactions of cosmic rays with the Earth's atmosphere. 

Given a probability map from \LIGO (Fig.~\ref{fig:ligomaps}), we define the cumulative coverage (\textit{coverage} in the following) at a given time from the trigger $t$ as the sum of the probabilities of all the pixels that have been observed by the LAT in the time window $[0,t]$:
\begin{equation}
C(t) = \sum_{h=0}^{H}~p_{h}~w_{h}(t),
\end{equation}
where $h=0...H$ are the \HEALPix pixels, and $w_{h}$ is 1 if the $h$-th pixel has been observed by the LAT between the trigger time and $t$, and 0 otherwise. We use the plot of $C(t)$ to decide the time window for the fixed time window search. An example for LVT151012 is shown in Fig.~\ref{fig:coverage}: the LAT was covering a region amounting to $\sim 50$\% of the total probability at the trigger time, and covered all the probability within $\sim 7$ ks. We then choose the time window $T_{\rm fix} = 0$--$8$ ks from the trigger time. Note that this is $\sim$1 ks longer than the exact time at which the coverage is equal to 1 in order to accrue some exposure in the regions of the sky that were covered last.

The analysis starts by selecting all pixels that were observable by the LAT during $T_{\rm fix}$ and within the 90\% containment of the \LIGO localization maps, down-scaled to a resolution (\textit{nside}$= 128$) of the order of the LAT Point Spread Function (PSF) at 100 MeV ($\sim$$12^\circ$). We then perform an independent likelihood analysis for each pixel, where we test for the presence of a new source at the center of the pixel. More specifically, for each pixel we consider a circular Region of Interest (RoI) with a radius of $8^{\circ}$ placed at the center of the pixel, and we consider all events detected during $T_{\rm fix}$ above 100 MeV. We then perform a standard unbinned likelihood analysis as described above. \red{This effectively amounts to a set of  likelihood-ratio tests. We detail in section ~\ref{sec:trials} how we correct for the multiple tests}. We then release to the community immediately all significant candidates (if any) through the GCN network\footnote{\texttt{https://gcn.gsfc.nasa.gov/lvc.html}}. Otherwise we proceed with the computation of the upper bound on the flux of the source, as described in the next section.

\subsection{Upper bound on the flux}

The scope of this part of the Fixed Time analysis is to provide one single constraint on the flux that accounts for the information on the localization of the event, and that can be applied in general to any model that predicts $\gamma$-ray counterparts to GW events.

Given the size of the LIGO error region, and the resulting number of RoIs we need to consider, the computation of an upper limit with the procedure described for example in \citet{2010ApJ...719..900K} is computationally prohibitive. Also, a frequentist framework does not provide a self-consistent way to account for the fact that different positions in the LIGO localization map have different probabilities. We hence adopt here a different approach, and consider the LIGO probability map as an informative prior for our analysis. Coherently, we adopt the Bayesian formalism, which also allows us to account in a natural way for uncertainties in our knowledge of the background (see below).

In order to avoid confusion, in this paper we distinguish between \textit{upper limit} and \textit{upper bound} on the flux of the EM counterpart, as suggested by \citet{2010ApJ...719..900K}. The upper limit at the $\alpha$ confidence level is defined as the largest flux for the source $S$ so that $S$ has a probability $1-\alpha$ of going undetected. As such, the upper limit is linked to the sensitivity of the instrument, the background level and the detection procedure. The upper bound, instead, is the upper edge of a (frequentist) confidence interval or a (Bayesian) credibility interval for the flux of the source, and hence depends on the observation at hand. According to this convention, in this paper we only deal with upper bounds.

For the Bayesian analysis described in this section, the difference between upper limit and upper bound is less pronounced than in the frequentist framework because in a Bayesian context detection and measurement are treated separately \citep{2011arXiv1103.2987R}. This is probably the reason why in the literature authors have called the upper bound of the credibility interval \textit{upper limit} \citep[for example]{2015PhRvD..91f2008A,Agashe:2014kda, 2007ASPC..371...75C, 1998PhRvD..57.3873F,2007ApJ...657.1026W}. However, in order to avoid confusion, we will call the upper edge of the Bayesian credibility interval \textit{upper bound} anyway in the rest of this paper.

The upper bound $F_{ub}$ on the photon flux $F$ of a source $S$ is the upper edge of a credibility interval starting at 0 and with a probability $p_{ub}$:
\begin{equation}
\int_{0}^{F_{ub}}~P(F|D)~dF = p_{ub}.
\label{eq:upper_limit_definition}
\end{equation}
Here $P(F|D)$ represents the posterior probability for a flux of $F$ given the data set $D$, marginalized over all the other parameters. Let us assume a power-law spectrum for the source $S$, with photon index $\alpha$ and photon flux $F$. We will compute upper bounds in a narrow energy band; thus this assumption does not impact the results much \citep{GW150914_LAT}. As explained above, the baseline model (background) is made up of the Galactic template, the isotropic template, and all sources from the 3FGL catalog. For simplicity, let us assume that this background model has no free parameters, i.e., we fix the normalization of the Galactic template and of the isotropic template to 1, and all the parameters for the 3FGL sources to their catalog values. We will relax this hypothesis later. Our likelihood model has then only 4 free parameters, namely $\alpha$, $F$, and the longitude and latitude pair $\vec{\delta}$ of $S$. 
The prior $\pi(\vec{\delta})$ on $\vec{\delta}$ is provided by the \LIGO probability map. 
Since we do not have any previous knowledge about what the flux of the source should be, we  use a prior $\pi(F)$ for $F$ uniform between a minimum of 0 and a maximum\footnote{The particular value chosen as maximum does not influence the final results, as long as it is much higher than the flux of a realistic source. For comparison, the brightest source ever detected by the LAT was GRB 130427A, that reached a photon flux of a few $10^{-3}$ photons cm$^{-2}$ s$^{-1}$.} of 100 photons cm$^{-2}$ s$^{-1}$ for $F$.
The use of a uniform prior in this context is suggested by the Particle Data Group \citep{Agashe:2014kda} and by other authors \citep{2015NIMPA.774..103B}, and it has been shown to produce results with good frequentist properties \citep{1983NIMPR.212..319H}. It is also desirable because the posterior probability obtained at the end of the procedure (Fig.~\ref{fig:lat_lvt151012}) can be re-weighted with other prior choices, should the reader have reasons to use other priors, as suggested by \citet{2015NIMPA.774..103B}. For similar reasons, we set a uniform prior $\pi(\alpha)$ between $-10$ and $10$ for $\alpha$ as well.  The posterior probability marginalized with respect to $\alpha$ and $\vec{\delta}$ can be written as:

\begin{equation}
P(F|D) = \int~\int~P(\alpha, \vec{\delta}, F| D)~d\alpha~d\Omega,
\end{equation} 
where $\Omega$ represents solid angle. Bayes' theorem allows us to write:
\begin{equation}
P(\alpha, \vec{\delta}, F| D) \propto P(D | \alpha, \vec{\delta}, F)~\pi(\alpha)~\pi(\vec{\delta})~\pi(F),
\label{eq:bayes_theorem}
\end{equation}
where $P(D | \alpha, \vec{\delta}, F)$ is the likelihood function for our data set $D$. We have also assumed that the prior $\pi(\alpha, \vec{\delta},F)$ can be written as $\pi(\alpha, \vec{\delta},F) = \pi(\alpha)~\pi(\vec{\delta})~\pi(F)$. 
In order to keep the marginalization computationally feasible, we substitute for the integration over the spatial dimension with a summation over the pixels of the \LIGO map, adopting the following approximation:

\begin{equation}
\int P(D | \alpha, \vec{\delta}, F) \pi(\vec{\delta}) d\Omega \simeq \Omega~\sum_{h=0}^{H} p_{h} P(D | \alpha, \vec{\delta}_h, F)
\end{equation}
where $p_{h}$ is the probability associated with the $h$-th pixel with center $\vec{\delta_{h}}$. We have also used the fact that \HEALPix is an equal-area projection, and we have called $\Omega$ the solid angle covered by each pixel. We can now rewrite eq.~\ref{eq:upper_limit_definition} as:

\begin{equation}
\Omega~\sum_{h=0}^{H} p_{h}~\int_{0}^{F_{ub}}~\int P(\alpha, \vec{\delta}_{h}, F | D)~d\alpha~dF = p_{ub}.
\end{equation}
In other words, the upper bound $F_{ub}$ is the value for which the following integral function:
\begin{equation}
U(x) = \Omega~\sum_{h=0}^{H} p_{h}~\int_{0}^{x}~\int P(\alpha, \vec{\delta}_{h}, F | D)~d\alpha~dF
\label{eq:ul_2}
\end{equation}
equals $p_{ub}$. 

In order to compute this integral in practice we use a Markov Chain Monte Carlo (MCMC) technique. However, eq.~\ref{eq:ul_2} involves a likelihood function that considers the entire dataset $D$ at once, \blue{and not the single RoIs independently, because it relates to the entire LIGO localization region.} It is therefore very expensive to compute. To reduce this complexity to a manageable level we start by observing that the MCMC can operate on the \textit{un-normalized} posterior, that is, on the product of the likelihood function and the prior that we wrote in eq.~\ref{eq:bayes_theorem}. In this analysis we use the \textit{unbinned} Poisson likelihood:

\begin{equation}
P(D|\alpha, \vec{\delta}_h, F) = \prod_{i=0}^{N}~m_{i}(\alpha, \vec{\delta}_h, F)e^{-m_{i}(\alpha, \vec{\delta}_h, F)},
\end{equation}
where $m_{i}$ is the photon density that the likelihood model yields at the energy and position of the i-th \textit{event}\footnote{An \textit{event} can be either a photon or a particle that has been mis-classified as a photon. Depending on the data class used in the analysis, the particle contamination can be more or less pronounced.}, and the product is performed over the \textit{events} in our selection. In reality, the model prediction $m_{i}$ does not depend on $\alpha$ or $F$ for all events that are separated from $\vec{\delta}_h$ by more than the typical size of the LAT PSF. This observation is particularly true in our case since the source is not detected because the flux $F$ is below the LAT sensitivity. In other words, if we define $m_{i}(\alpha, \vec{\delta}_h, F) = m_{\rm bkg} + m_{\rm src}(\alpha, \vec{\delta}_h, F)$, the contribution from the source $m_{\rm src} \sim 0$ if we are far enough from $\vec{\delta}_h$. Let us define an RoI centered on $\vec{\delta_{h}}$ and a radius big enough so that outside of the RoI $m_{\rm src} \sim 0$. We can then write:

\begin{equation}
P(D|\alpha, \vec{\delta}_h, F) \propto \prod_{D_{\rm RoI}}~m_{i}(\alpha, \vec{\delta}_h, F)e^{-m_{i}(\alpha, \vec{\delta}_h, F)},
\end{equation}
where here the product is performed only over the events within our RoI ($D_{\rm RoI})$. We can now write Bayes' theorem for one RoI as:
\begin{equation}
P(\alpha, \vec{\delta}_h,F|D) \propto P(D_{\rm RoI} | \alpha, \vec{\delta}_h, F)\pi(\alpha)\pi(F)
\end{equation}
This consideration allows us to compute the integral in eq.~\ref{eq:ul_2} by collecting separately (and in parallel) $n_{s}$ samples from each RoI centered around each of the $\vec{\delta}_{h}$, and then merging these samples in one set weighted according to the probability of the corresponding pixel $p_{h}$. We can then base our inference on this merged set as representative of the full posterior probability.

In order to account for uncertainties in the background model we now introduce two more parameters for each RoI, i.e., the normalizations of the isotropic template $I$ and of the Galactic template $G$. We use a uniform prior $\pi(I)$ between 0 and 100 for $I$. For $G$ instead we adopt a Gaussian prior with average 1 and standard deviation 0.15, as a conservative estimate of our systematic uncertainty on this component. Since we introduce these two parameters in each one of the $n_{\rm RoI}$ RoIs separately, they amount to $2 \times n_{\rm RoI}$ nuisance parameters that are marginalized out in the final posterior. This allows us to take into account likely differences in background uncertainties for different regions of the sky. \red{Similarly, we also free the normalizations of all variable 3FGL sources that are flaring during the interval considered, as determined by the standard monitoring procedures of the \Fermi/LAT collaboration \citep{2013arXiv1303.4054C, 2013ApJ...771...57A}. Also in this case these nuisance parameters are marginalized over in the final posterior. The typical number of sources in a flaring state within the typical LIGO region is $\lesssim 1$ per day, and the difference between the average state and the flaring state can be detected on the time scales of this analysis ($\simeq 10$ ks) only for extremely bright flares. Typically the effect of a flaring source on its RoI is also smoothed out when marginalizing over the many RoIs defined from the LIGO map. Thus, the effect of this correction will be noticeable only if an exceptionally bright flaring source happens to be in a region of high probability of the LIGO map.}

In practice, for each RoI we collect $n_{\rm S}=50,000$ samples $\vec{q}_{s} = (\alpha_s, F_s, I_s, G_s)$, with $s=1..n_{\rm S}$ using the sampler \textit{emcee}\footnote{The use of more powerful samplers such as MultiNest \citep{2009MNRAS.398.1601F} is unnecessary in this case, because we have only 4 parameters and we have verified that the posterior is not multimodal. Moreover, we are not interested in the Bayes' factor that cannot be directly computed given our approximations.} \citep{2013PASP..125..306F}. For each sample, we also compute the corresponding \textit{energy} flux $f_{s}$ by integrating the power-law with parameters $F_{s}, \alpha_{s}$. We then estimate the integral of the marginalized posterior distribution for $F$ and $f$ by building the integral distribution of the $F_{s}$ and the $f_{s}$ and then using interpolation to obtain $u_{h}(x) = \int_{0}^{x}~\int~P(\vec{q}, \vec{\delta}_{h}, F | D)~d\vec{q}~dF$ (or the equivalent for $f$). 
The number of pixels outside of the 90\% containment is much greater than within it but they do not contribute significantly to the probability (the sum of their probabilities is 0.1 by definition). Since considering all pixels would be computationally prohibitive, we instead assume that they have on average a posterior similar to the average of the pixels within the 90\% containment. This is a reasonable assumption because even if the exposure of the LAT over the considered time scales is not perfectly uniform, all these points have low probabilities, so that any local variation is smoothed out. We have then $U(x) = \sum_{h=0}^{H} p_{h}~u(x)$.

In principle the function $U(x)$ can be used to estimate the upper bound for any probability $p_{ub}$. However, the posterior is completely determined by the choice of the prior for $F$ \textit{when $F$ is small}, i.e., the prior is no longer ``uninformative''. Indeed, as $F$ becomes smaller and smaller, the predicted number of counts from the source is $~\sim 0$ and the likelihood becomes constant and independent of $F$ and $\alpha$. For this reason, $p_{ub}$ should be chosen so that the likelihood contains information about the corresponding flux level. As an example, we report in Figure ~\ref{fig:lat_lvt151012} the posterior probability for the energy flux in the 0.1 -- 1 GeV energy range for LVT151012. The flux level at which the blue curve in the right panel intersects a given probability $p_{ub} = P(< F)$ corresponds to the upper bounds at that credibility level. In the left panel we can see that the posterior probability clearly deviates from a flat curve much earlier than the fluxes corresponding to the 90\%, 95\% and 99\% credibility levels, as the likelihood function dominates over the prior at those fluxes.
\begin{figure}[tb]
\centering
\includegraphics[width=0.4\textwidth,trim=20 0 0 0,clip=true]{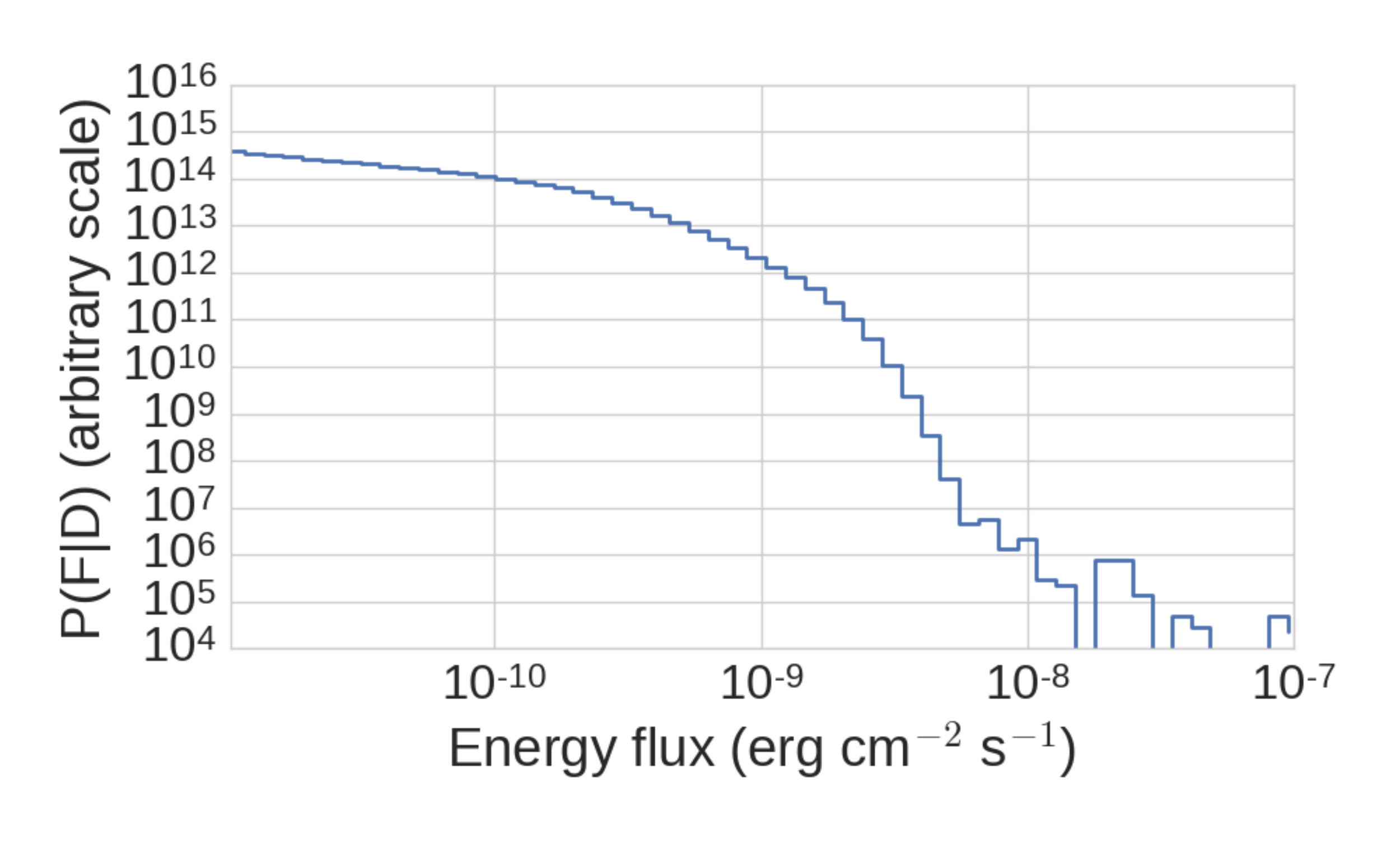}
\includegraphics[width=0.4\textwidth,trim=20 0 0 0,clip=true]{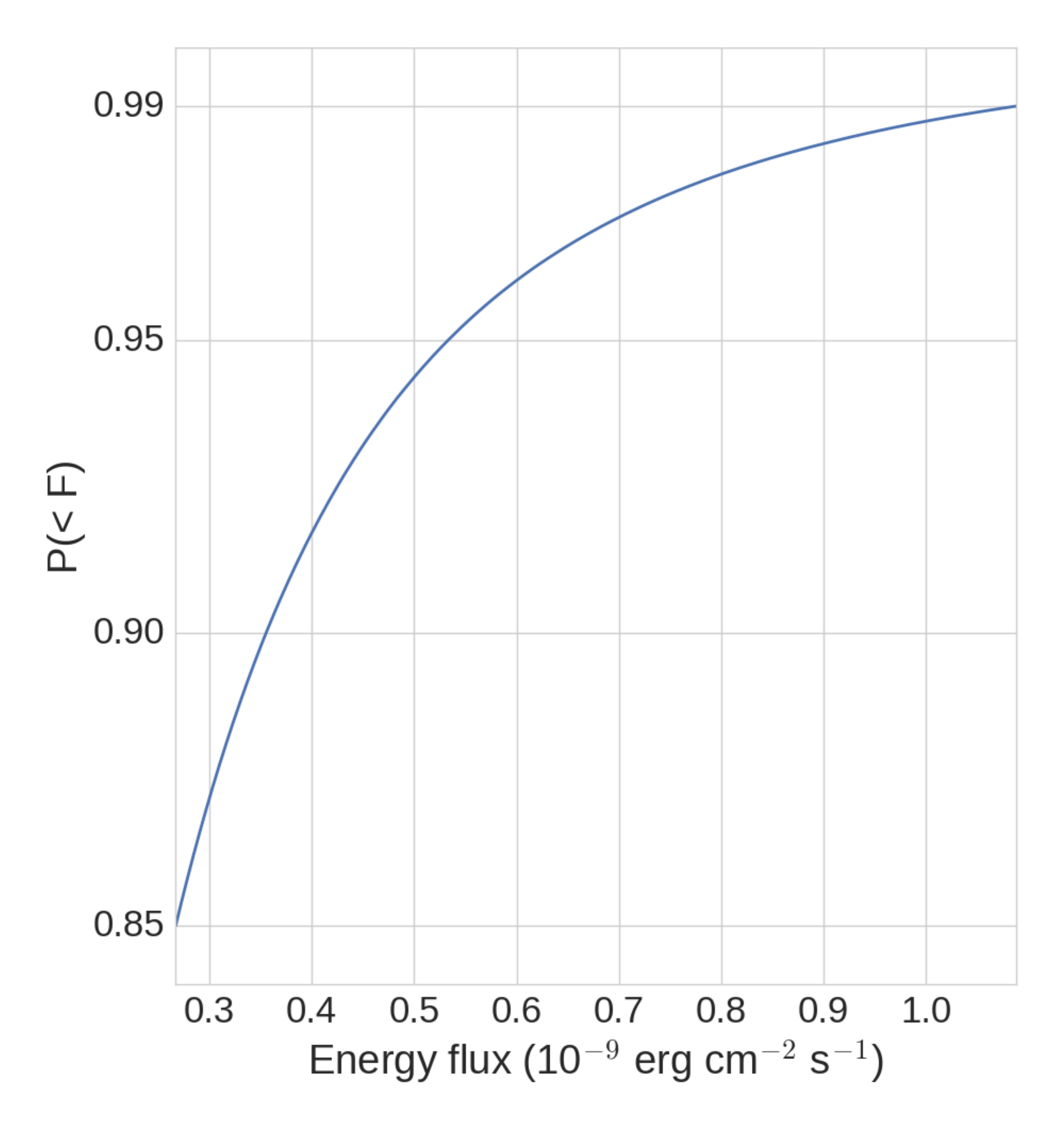}
\caption{Differential (\blue{top}) and integral (\blue{bottom}) marginalized posterior for the 0.1--1 GeV energy flux for LVT151012 and the time window $0$--$8$ ks. In the \blue{top} panel, we note that the rightmost part of the distribution is affected by sampling noise. In the \blue{bottom} panel \citep[reproduced from][]{Racusin16}, the flux at which the blue curve intersects a given probability $P(F < x)$ corresponds to the upper bound at that credibility level.}
\label{fig:lat_lvt151012}
\end{figure}

\subsection{Adaptive time window search}
\label{sec:adaptive-time-search}
We now describe an alternative analysis, which maximizes the time window for each point in the sky separately in order to get the largest possible exposure close to the trigger time. In case of a non-detection, it also provides a map of upper bounds on the flux. \blue{Note that in this analysis, we do not combine the values of the flux upper bounds of all the pixels in one global measurement, as we do in the fixed time window analysis.} The map is useful if more accurate information on the localization of a possible counterpart, for example from its detection by some other instrument, becomes available after the analysis has been performed and published. 
In this case, depending on the position of such a localization, the reader can choose the upper bounds most relevant for the candidate counterpart \textit{a posteriori}. \blue{Because the time windows have been optimized for each pixel, the corresponding bounds could be deeper and hence more constraining with respect to the upper bound provided by the fixed time window analysis described in the previous section.}

As in the previous analysis we perform an independent likelihood analysis for each pixel in the \LIGO map, where we test for the presence of a new source at the center of the pixel. However, here we optimize separately the time window for each pixel. We define 8$^\circ$ radius RoIs centered at the center of each pixel.
For the $h$-th pixel we define the interval $T_{h}^{\rm ad}$ that starts when the center $\vec{p}_{h}$ of the RoI becomes observable by the LAT, i.e., when it is $<$ 65$^\circ$ from the LAT boresight and $<$ 92$^\circ$ from the local zenith, and ends when the $\vec{p}_{h}$ is no longer observable\footnote{Note that, considering that the size of the RoI is 8$^\circ$, the 92$^\circ$ zenith cut is equivalent to requiring that every point of the RoI has a zenith $ < 100^\circ$.}.
First we measure the test-statistic value obtained for all the RoIs. Then we apply the method described in section~\ref{sec:trials} to asses whether we have one or more detections of new sources. If we do not find any significant excess, we compute a flux upper bound for each pixel separately using the technique described in \citet{1983NIMPR.212..319H}, where we profile out all nuisance parameters.

Figure \ref{fig:lat_ul_151012_adap} shows an example of the results returned by this analysis for LVT151012. \blue{For a given location within the LIGO probability contour, the top panel can be used to extract the value of the flux upper bound. From the second panel, instead, the time when the RoI entered the LAT FoV can be extracted. Since the color of the pixels in this panel, matches the color of the horizontal bars in the third panel, the interval $T^{\rm ad}_{h}$ for that particular position can be derived.}

\begin{figure*}
\centering
\includegraphics[width=0.65\textwidth,trim=100 0 100 25,clip=true]{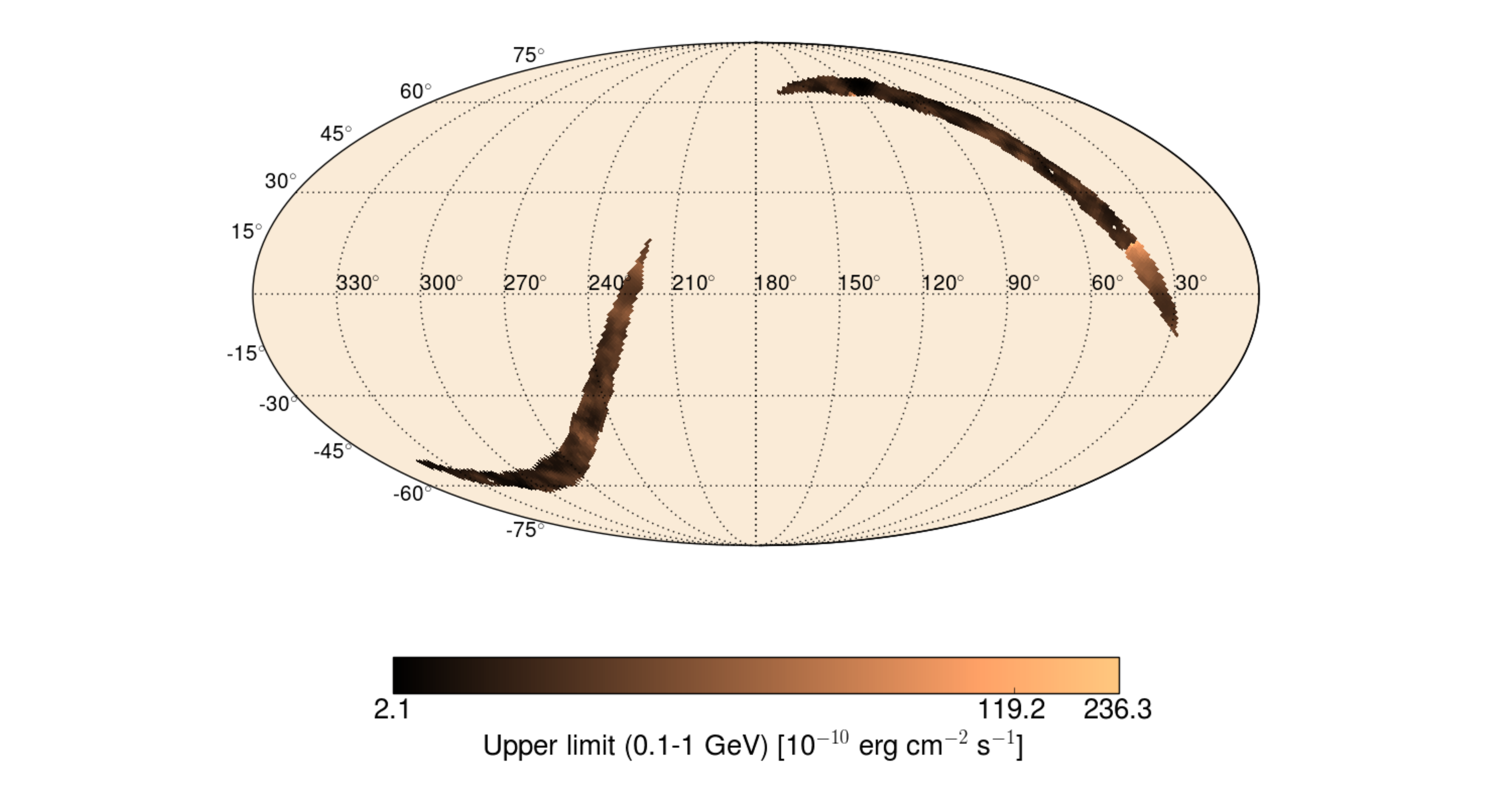}
\includegraphics[width=0.55\textwidth,trim=0 30 30 10,clip=true]{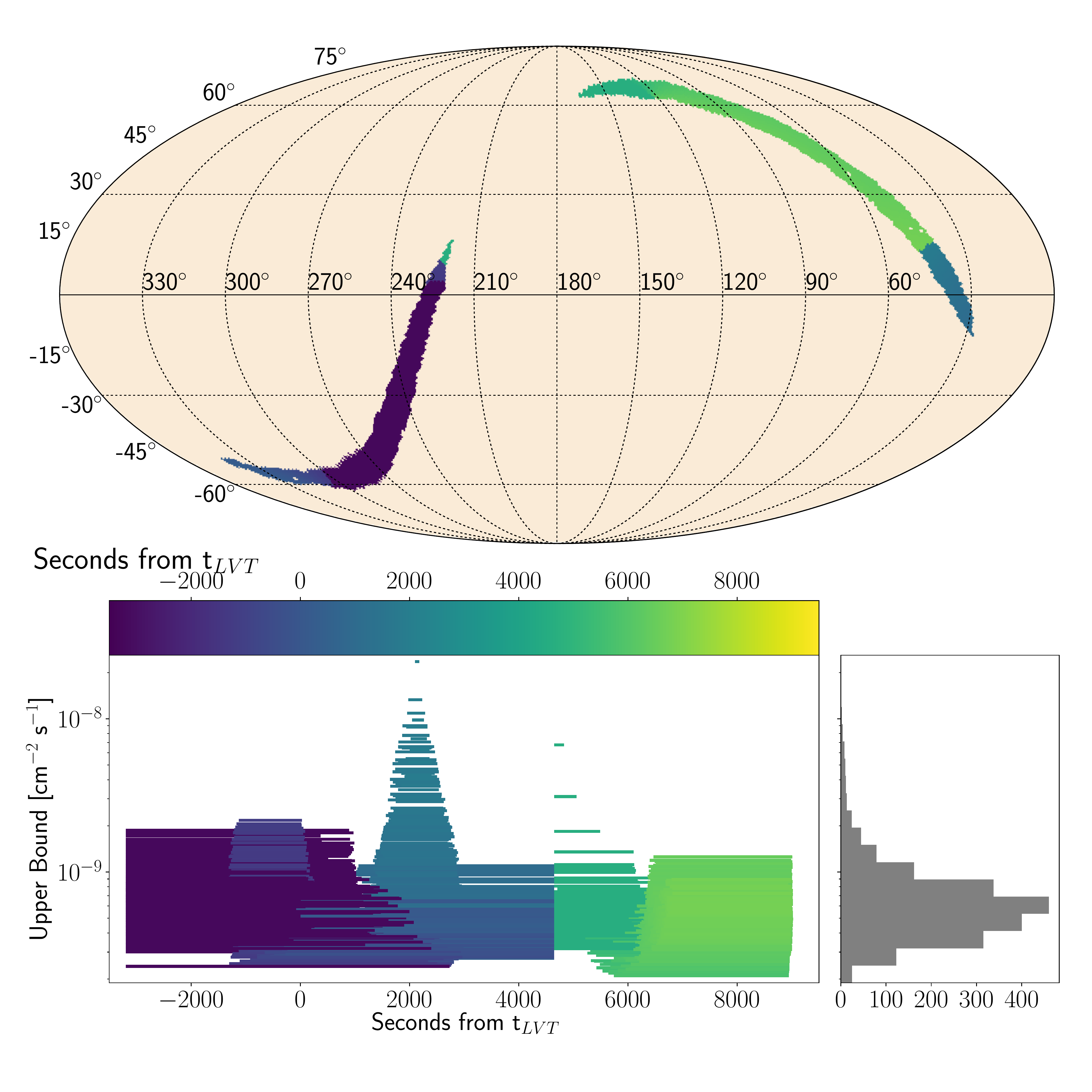}
\caption{The adaptive time interval analysis for LVT151012 over the first \Fermi orbit containing $t_{LVT}$: flux upper bound map during $T_{h}^{\rm ad}$ ({\it top}), the start time for $T_{h}^{\rm ad}$ relative to $t_{LVT}$ for each pixel ({\it middle}), and the upper bound with the corresponding time window for each RoI ({\it bottom}).
The vertical position in the bottom plot corresponds to the value of the LAT upper bound, and the horizontal line marks $T_{h}^{\rm ad}$. 
The color of each line indicates the time when the RoI entered the LAT FoV, and matches exactly the color of the pixel in the middle panel. The horizontal histogram displays the distribution of the upper bounds. Reproduced from \citet{Racusin16}.}
\label{fig:lat_ul_151012_adap}
\end{figure*}

\subsection{False discovery rate}
\label{sec:trials}

\red{In both analyses presented in the previous sections we are considering multiple RoIs and testing for the presence of sources at different locations, i.e., we are performing $m$ multiple likelihood-ratio tests. In principle we can control the family-wise error rate (FWER), defined as the probability of rejecting the null-hypothesis when it is true (Type I error) at least once in the set of $m$ tests, by setting an appropriate threshold ${\rm TS}_{\rm thr}$ in TS corresponding to, say, a $5\sigma$ rejection criterion. Because of the multiple tests, ${\rm TS}_{\rm thr}$ will be larger than 25, which is the value corresponding to $5\sigma$ for one test as explained above. However, given the correlation between different RoIs, and the complexity of the procedure, we do not have any theoretical expectations on ${\rm TS}_{\rm thr}$. We could estimate it by repeating our analysis many times on Monte Carlo simulations and directly measuring the distribution of ${\rm TS}$ under the null hypothesis. This is not feasible given the computational resources required for a single analysis. Instead, we choose a procedure that controls the False Discovery Rate (FDR) instead of the FWER, i.e., the proportion $\alpha$ of false detections on a set of detections. This allows the method to have good statistical power while maintaining control on Type I errors, and it is a much more tractable statistical problem. The original method to control the FDR was proposed by \citet{benjamini1995controlling}, and it is valid for independent tests. In our case, however, neighboring pixels are correlated because of the PSF. We therefore adopt the correction described in \citet{hopkins2002new}, so our procedure becomes:
\begin{enumerate}
\item Order the p-values $P_i$ of all $N$ pixels within the LIGO 90\% containment region from the smallest to the largest to build the set ${P_{1}, ..., P_{N}}$. Since the TS for a single pixel is distributed as $\frac{1}{2}~\chi^2$ with 1 d.o.f. \citep{1996ApJ...461..396M}, the p-value of the $i$-th pixel is computed as:
\begin{equation}
P_{i} = 1/2 ~ \int_{{\rm TS}_{i}}^{\infty}\chi^{2}(x)~dx
\end{equation}
\item Find the maximum index $k$ where the condition $P_i \leq i~\alpha / (N C_{N})$ holds true, where the factor $C_{N} = \sum_{j=1}^{n} 1 / j$ corrects for the correlation between $n$ neighboring pixels \citep{hopkins2002new}. We consider as correlated all pixels within a PSF. The maximum PSF size in the energy range used for our analysis is $R \sim 12^{\circ}$ at 100 MeV (95\% containment radius). Hence, we use for $n$ the number of pixels in the HEALPix map contained within a disk of radius $R$\footnote{The typical HEALPix map used in our analysis has \texttt{nside}=128, and we obtain $n=2244$ pixels.}
\item Reject the null hypotheses whose p-values are less than or equal to $P_{k}$ in ${P_{i}}$, i.e., consider as detections all pixels with $0 \le i \le k$.
\end{enumerate}
We set an FDR of $\alpha = 0.01$. We stress again that $\alpha$ is not the FWER, i.e., the probability of obtaining one false detection, but the proportion of Type I errors among all detections. Therefore, despite potential appearances, it is a conservative threshold because it means that on average we will detect 100 new sources before committing a Type I error. For all candidates, we will then assess the significance by performing Monte Carlo simulations. In order to reduce the number of required simulations to a manageable level, we will adopt the method described in \citet{2011APh....35..230V} that allows an estimate of the significance with just a few ($\sim$10) simulations.}


The procedure we just detailed requires the null hypothesis to be a good representation of the case with no source, and that our Monte Carlo procedure faithfully simulates a real observation. To verify this, we compared the distribution of observed test-statistic values in one instance of our analysis for LVT151012 with the expected distribution obtained from a Monte Carlo simulation. We generated a full-sky simulation using the \texttt{gtobssim} tool. The sky model consists of our null hypothesis, i.e., the Galactic emission template (we used the standard Galactic template  \texttt{gll\_iem\_v06.fits}), the isotropic diffuse emission (tabulated in the \texttt{iso\_P8R2\_TRANSIENT010E\_V6\_v06.txt}) and all the sources from the 3FGL catalog \citep{3fgl}. The actual pointing history of the satellite during the \LIGO event was used so that the correct exposure of the sky was automatically taken into account. Then, we repeated the adaptive time window search on the simulated data. In Fig.~\ref{fig:MC} we compare the TS distribution obtained from flight data and Monte Carlo for LVT151012.
The distribution obtained from simulated data is a good match to the distribution of the TS values obtained from real data, and the good absolute agreement confirms that the model used in the simulation is a good representation of the sky and that the systematics of the analysis are well under control. As a result, our simulation procedure is suitable for computing the distribution of TS under the null hypothesis that no transient signal is present.

\begin{figure}[tb]
\centering
\includegraphics[width=0.48\textwidth]{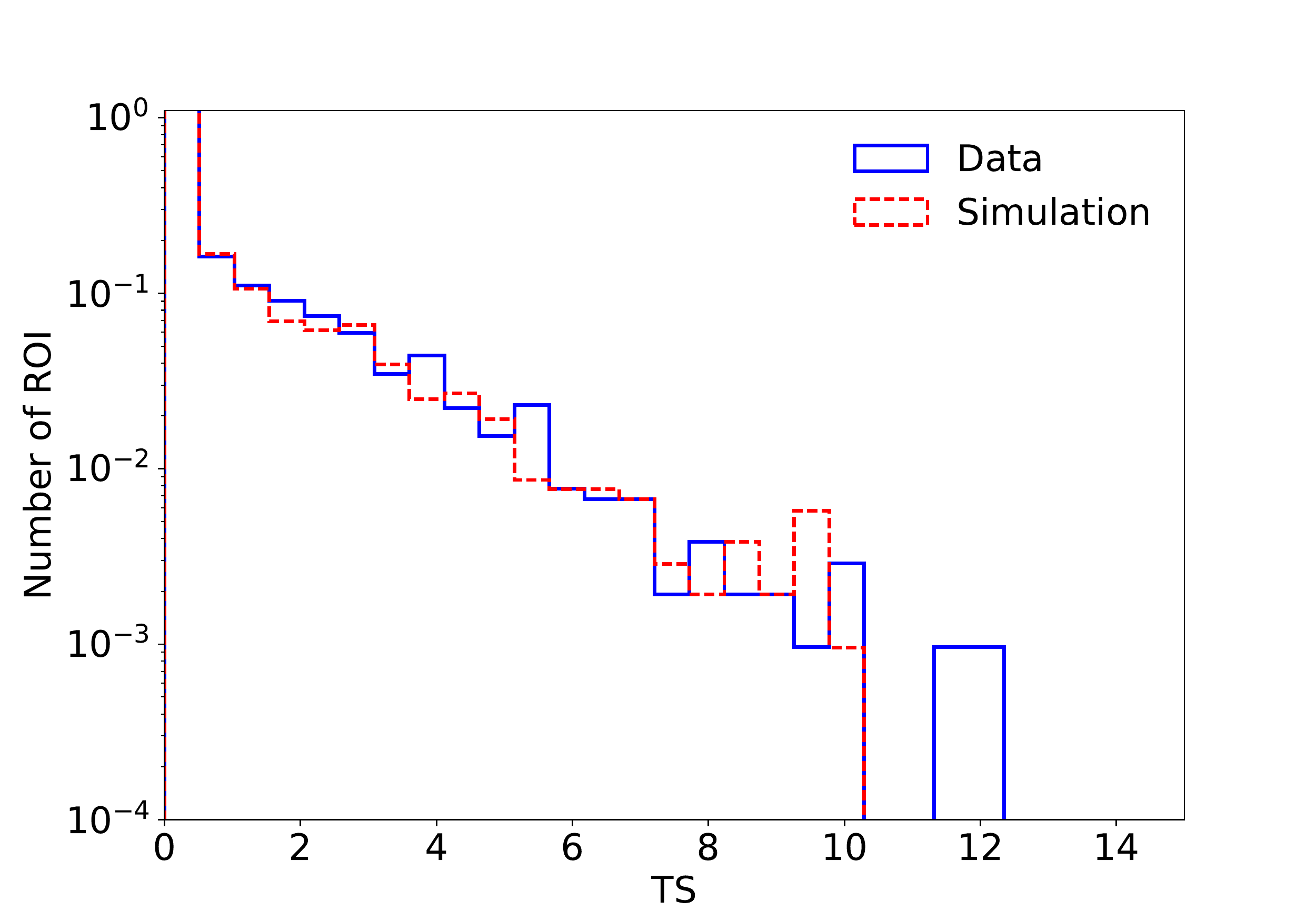}
\caption{Comparison between the TS distributions of the data (solid blue histogram) and Monte Carlo simulations (dashed red \blue{histogram} ). 
} \label{fig:MC}
\end{figure}

\section{Summary and conclusion}\label{sec:disc}

With its wide FoV and its survey capabilities, \Fermi-LAT is suitable for looking for EM counterparts to GW events above 100\,MeV and for constraining their fluxes. In particular, in cases of NS-NS or NS-BH mergers, the expected EM counterpart is an sGRB. The LAT is a wide-field instrument that routinely detects and localizes GRBs during their afterglow phase. If the position measured by other instruments such as \Fermi-GBM during the prompt phase has an uncertainty too large for follow up by X-ray, optical and radio telescopes, a LAT detection and localization can vastly improve the chances for a successful follow up. Moreover, the LAT is one of few instruments that can constrain the flux from the EM counterpart despite a large uncertainty on its position.

We have presented two novel techniques to perform the search for an EM counterpart to a GW event in \Fermi-LAT data and to provide constraints on its flux. They fully exploit at the same time the capabilities of the instrument as well as the prior information available from the \LIGO/\VIRGO observatories. 
These methods, developed during the first \LIGO science run `O1', will be systematically used to search for EM counterparts to future GW events. In case of a detection, the methods presented here will return a localization, a flux estimation and a significance of the EM counterpart. If no EM counterpart is detected, a meaningful set of constraints on the flux of the source can be measured.
\blue{The methods are now implemented in an automatic analysis pipeline triggered by \LIGO/\VIRGO. This minimizes the turn-around time and increases the chance, in case of a detection, of initiating a prompt follow-up campaign to detect the EM counterpart at other wavelengths.}

The authors thank A.~Strong (\blue{Max-Planck institute f\"ur extraterrestrische Physik}), J.~Conrad (Stockholm University) and N.M.~Mazziotta (\blue{INFN Sezione di Bari}) for the helpful discussion on the STAT mailing list of the \Fermi-LAT collaboration.

Some of the results in this paper have been derived using the HEALPix \citep{HEALPix} package.

The \Fermi-LAT Collaboration acknowledges generous ongoing support from a number of agencies and institutes that have supported both the development and the operation of the LAT as well as scientific data analysis. These include the National Aeronautics and Space Administration and the Department of Energy in the United States, the Commissariat \`a l'Energie Atomique and the Centre National de la Recherche Scientifique / Institut National de Physique Nucl\'eaire et de Physique des Particules in France, the Agenzia Spaziale Italiana and the Istituto Nazionale di Fisica Nucleare in Italy, the Ministry of Education, Culture, Sports, Science and Technology (MEXT), High Energy Accelerator Research Organization (KEK) and Japan Aerospace Exploration Agency (JAXA) in Japan, and the K.~A.~Wallenberg Foundation, the Swedish Research Council and the Swedish National Space Board in Sweden.
 
Additional support for science analysis during the operations phase is gratefully acknowledged from the Istituto Nazionale di Astrofisica in Italy and the Centre National d'\'Etudes Spatiales in France.

\bibliographystyle{yahapj}
\bibliography{references}


\end{document}